\documentclass[aps,prd,floatfix,superscriptaddress,preprintnumbers,nofootinbib,preprint,tightenlines,longbibliography]{revtex4-1}
\pdfoutput=1

\usepackage{amsmath, amssymb, amsfonts, graphicx, mathrsfs, verbatim, float, slashed, bbm, color, xcolor, xspace, enumerate, url, bbding, multirow, outlines,upgreek}
\usepackage[english]{babel}
\usepackage[colorlinks=true,linkcolor=black,citecolor=galacticcenterbubblegum]{hyperref}
\definecolor{galacticcenterbubblegum}{rgb}{0.8,0, 0.8}

\begin{document}

\title{Galactic Center Gas Clouds and Novel Bounds on Ultra-Light Dark Photon, Vector Portal, Strongly Interacting, Composite, and Super-Heavy Dark Matter}

\author{Amit Bhoonah}
\affiliation{The McDonald Institute and Department of Physics, Engineering Physics, and Astronomy, Queen's University, Kingston, Ontario, K7L 2S8, Canada}

\author{Joseph Bramante}
\affiliation{The McDonald Institute and Department of Physics, Engineering Physics, and Astronomy, Queen's University, Kingston, Ontario, K7L 2S8, Canada}
\affiliation{Perimeter Institute for Theoretical Physics, Waterloo, Ontario, N2L 2Y5, Canada}

\author{Fatemeh Elahi}
\affiliation{School of Particles and Accelerators, Institute for Research in Fundamental Sciences IPM, Tehran, Iran}

\author{Sarah Schon}
\affiliation{The McDonald Institute and Department of Physics, Engineering Physics, and Astronomy, Queen's University, Kingston, Ontario, K7L 2S8, Canada}
\affiliation{Perimeter Institute for Theoretical Physics, Waterloo, Ontario, N2L 2Y5, Canada}

\begin{abstract}
Cold gas clouds recently discovered hundreds of parsecs from the center of the Milky Way Galaxy have the potential to detect dark matter. With a detailed treatment of gas cloud microphysical interactions, we determine Galactic Center gas cloud temperatures, unbound electron abundances, atomic ionization fractions, heating rates, cooling rates, and find how these quantities vary with metallicity. Considering a number of different dark sector heating mechanisms, we set new bounds on ultra-light dark photon dark matter for masses $10^{-22}-10^{-10}$ eV, vector portal dark matter coupled through a sub-MeV mass boson, and up to $10^{60}$ GeV mass dark matter that interacts with baryons. 
\end{abstract}

\maketitle

\tableofcontents


\section{Introduction}
\label{sec:intro}

Although its gravitational influence has been observed in galactic and cosmological dynamics, dark matter's non-gravitational couplings and cosmological origin remain unknown. Discovering dark matter's features would increase our knowledge of the structure and history of the physical universe. In this work we determine how cold gas clouds near the center of the Milky Way galaxy can be used to detect dark matter.

Some prior work has considered the impact dark matter can have on interstellar and intergalactic gas \cite{Chivukula:1989cc,Qin:2001hh,Bhoonah:2018wmw}. The physical basis of these studies is remarkably simple: dark matter tends to have a higher temperature than the coldest interstellar and intergalactic gas. Therefore dark matter, which is often more dense than diffuse astrophysical gas, can heat this gas to higher-than-observed temperatures, if dark matter interacts enough with baryons or electrons in the gas cloud. A somewhat distinct mechanism for dark matter heating of interstellar gas was more recently identified in \cite{Dubovsky:2015cca}. A long-range and slowly shifting electric potential will heat interstellar gas by transforming electromagnetic potential energy into gas kinetic energy, through acceleration and collision of gas particles. Such a long-range oscillating electric potential is sourced by vector-like dark matter with a very long compton wavelength and a small mixing with the photon of the Standard Model, known as ultra-light dark photon dark matter. Most recently, in reference \cite{Bhoonah:2018wmw} the authors of this paper identified that cold gas clouds near the Galactic Center provide unprecedented sensitivity to dark matter capable of heating interstellar gas. In the following work, the best bounds are set by the coldest cloud, denoted G1.4-1.8+87. We want to here highlight the concern that for this particular object the temperature measurement is the result of a single-channel fluctuation and therefore the true temperature maybe considerably hotter than $22$ K. The cloud is nonetheless included, albeit marked as 'preliminary' to indicate its status. We strongly encourage follow up observations to confirm relevant properties, as well as to motivate future searches for these types of cold gas clouds. For further discussion see Section \ref{sec:add}.

A few hundred cold gas clouds were recently discovered, each situated a few hundred parsecs from the center of the Milky Way Galaxy \cite{McClure-Griffiths:2013awa}. Based on their speed, it seems likely that these gas clouds were formed tens of millions of years ago in the central molecular zone, before being pushed by a galactic wind out of the core of the galaxy. These gas clouds provide un-paralleled sensitivity to dark matter with certain interactions, as explored in \cite{Bhoonah:2018wmw}. Specifically, cold gas clouds provide a physical environment unobtainable by terrestrial experiments: cold gas clouds contain hundreds of solar masses of ionized gas, which is rather sensitive to heating by dark matter with relatively strong couplings to baryons, or dark matter with a long-range interaction with electrons. One example of the latter is dark matter with a small electromagnetic charge, often called milli-charged dark matter. In \cite{Bhoonah:2018wmw}, cold Galactic Center gas clouds placed leading constraints on dark matter strongly coupled to baryons and millicharged dark matter.

Here we study the properties of Galactic Center gas clouds in more detail using a numerical code, and derive additional bounds on dark matter models. The remainder of this paper proceeds as follows.  In Section \ref{sec:gcprop} we review gas cloud physics and study the properties of gas clouds discovered near the Galactic Center, modeling their thermal properties using the numerical code CLOUDY. A new bound on ultra-light dark photon dark matter is obtained in Section \ref{sec:ldph}. The sensitivity of Galactic Center gas clouds to vector portal dark matter coupled through a sub-keV mass mediator is shown in Section \ref{sec:subMevmed}. New bounds on dark matter which interacts with baryons through a spin-independent coupling are found in Section \ref{sec:strong} along with the first derivation of the gas cloud ``overburden"; it is found that gas cloud bounds apply to dark matter as massive as $\sim 10^{60}$ GeV. In Section \ref{sec:disc}, we conclude with some discussion of how the temperature profile of gas clouds might be used to discover dark matter.

\section{Galactic Center Gas Cloud Properties}
\label{sec:gcprop}
In a previous paper \cite{Bhoonah:2018wmw} we showed that cold, atomic gas clouds near the Galactic Center provide a unique testing ground for a number of dark matter models. That work utilized simple volumetric interstellar gas cooling rates we obtained from \cite{DeRijcke:2013zha}. However, the physics governing the thermal, ionization and chemical state of this inter-galactic gas is complex. Here we initiate a fuller treatment of Galactic Center gas cloud dynamics. Specifically, we make use of the gas microphysics code CLOUDY, last described in \cite{2017RMxAA}, to simulate objects matching the physical properties of the clouds observed by McClure-Griffiths et al \cite{McClure-Griffiths:2013awa} used in our prior analysis. These simulations yield improved ionization and cooling rates, which are essential to accurately determine cold gas cloud sensitivity to dark matter models.

\subsection{ISM and Gas Physics}
\label{subsec:gasphysics}

The study of low-density interstellar medium (ISM) atomic gas dynamics is an active area of research. Understanding the composition and thermal properties of the ISM is critical to understanding the formation and evolution of galaxies and stars. There are a number of processes that play key roles in governing the state and evolution of the ISM. For our purposes we will be focusing on the factors regulating the heating and cooling of the gas. We will particularly investigate how the composition of the gas clouds and variation of the incident radiation fields change the thermal balance within the cloud and subsequently the ionization fraction and cooling rates.

The bulk of the ISM is made of hydrogen and helium. As the stellar population matures, the gas is further enriched with metals (here taken in the astrophysical context to mean any nuclei heavier than helium) as well as grains and more complex molecules. These additional components facilitate some of the dominant heating and cooling mechanisms, while also contributing to the regulation of the chemical network, which has further consequences for the gas's internal thermal processes. 

Metallicities within the Milky Way's ISM are determined in a number of ways.  Optical emission lines from singly and doubly ionized oxygen, OII and OIII, can be used to constrain oxygen abundances within Galactic HII (ionized hydrogen) regions. Then total metallicities are inferred assuming linear scaling between the two \cite{2000MNRAS.311..329D}. Using a similar method, observations of UV absorption lines can be used to constrain the abundance of a variety of elements and subsequently the total metallicity \cite{1996ARA&A..34..279S}. The spectra of stars can also constrain metallicities at the time of the star's formation and the combination of these observations shows the existence of a large scale metallicity gradient across the Galaxy \cite{2000A&A...363..537R,2007yCat..21620346R}. 

Grains (or dust) \cite{1995A&ARv...6..271D} encompass material ranging from simple molecules to complex particles up to $\sim 0.3 \mu\mathrm{m}$ in size. The grain population is split almost equally between silicate minerals and carbonaceous material. While the dust component only makes up a tiny fraction of the total ISM mass, it plays a significant role through its absorption and scattering of light, as a tracer of underlying physical conditions and perhaps most significantly through its direct interaction with the rest of the ISM, affects the overall chemical composition and subsequently star formation. Dust grain interactions with the gas include \cite{1992dge..book.....W} heating through photoelectric emission, formation of $\mathrm{H}_{2}$ molecules from emission off of grains, and coupling magnetic fields to the neutral gas.

The thermal balance of the ISM is determined by the heating sources and the cooling mechanisms available to the gas. Energy is injected from various background radiation fields including the UV background, cosmic ray background as well nearby stellar objects. Important heating mechanisms include \cite{doi:10.1146/annurev-astro-082708-101823} heating by low energy cosmic rays, photoelectric heating by grains, photoelectric heating by photoionization of atoms and molecules, grain-gas thermal exchange as well as hydrodynamic/magnetohydrodynamics heating and interstellar shocks, though the latter two are not included in this treatment. The most important processes for the calculation performed in this work are heating from cosmic rays and grains. Depending on the make-up of the gas, cooling can be facilitated by \cite{2010PhR...495...33B} metal line transitions, collisional ionization, Ly$\alpha$ photons, recombination, bremsstrahlung and molecular cooling. For the low density environment found in the gas clouds at hand, the most relevant processes are collisional excitation and subsequent decay of various metal species. In particular, carbon, oxygen and iron provide the dominant cooling terms at the temperatures of interest here \cite{2013MNRAS.433.3005D,doi:10.1146/annurev.aa.10.090172.002111,1995ApJ...443..152W}.
   
Energy is injected into the cloud via a radiation background. For this calculation we take into account both the cosmic UV/X-ray background \cite{doi:10.1146/annurev.aa.29.090191.000423,2009ApJ...703.1416F} as well as the cosmic ray background \cite{1742-6596-47-1-002,2006Sci...314..439A,1742-6596-47-1-002}. Given these sources of energy in combination with the relevant heating and cooling mechanisms, different ISM regions are often labelled according to their thermal and chemical phases \cite{1969ApJ...155L.149F,1977ApJ...218..148M}. These phases are typically called the cold neutral, warm neutral and hot ionized medium. Additional names for phases include the warm ionized medium \cite{2009RvMP...81..969H} and cold molecular phase \cite{2007prpl.conf...81B}. For a further review of the topic, see \cite{RevModPhys.73.1031}. However, factors such as turbulence, inflow of ga,s and supernova feedback can disrupt these distinct phases and complicate this picture of the ISM considerably \cite{refId0}. Therefore, we will not attempt to label the thermal and chemical phases of our cold gas clouds, although for the most part, the gas clouds we study here would fall in the ``cold neutral" category.

\subsection{CLOUDY Models}
\label{subsec:CLOUDY}

The CLOUDY code works by calculating equations of energy, mass and charge conservation, in addition to the detail balance equation governing the density of (ionized) atomic species $n_i$:
\begin{equation}
\frac{\partial n_{i}}{\partial t} = \sum_{j \neq i} n_{j} R_{ji} + {\rm source} - n_i(\sum_{ji}R_{ij} + {\rm sink}) = 0 \, [\mathrm{cm}^{-3}\mathrm{s}^{-1}],
\end{equation}
where $R_{ij}$ is the total rate at which species $i$ goes to $j$. Processes such as photo- and collisional ionization, recombinations and charge exchange contribute to the ``source" and ``sink" portions of the above equation. The relevant references for the physical processes can be found in \cite{2017RMxAA} and \cite{2013RMxAA..49..137F} as well as the CLOUDY documentation.
The additional parameters specified in CLOUDY for each gas cloud model are listed below:
\begin{itemize}

\item Metallicity - We assume the metallicity of the cloud scales with solar metallicity -- although the numerical code we use does account for depletion of individual atomic species due to absorption onto dust grains. While the metallicity gradient across the disk of the Milky Way is relatively well documented, the environment within the Galactic Center is perhaps not as straightforward. References \cite{0004-637X-809-2-143} and \cite{2016PASA...33...22N} both find stars with a range of metallicities within the Galactic bulge, which is indicative of both older, metal poor as well as newer populations of stars being present. As we are here working under the assumption that our population of gas clouds originates from within this Galactic Center region, we will investigate both metal rich and poor in our gas cloud models.

\item Dust Grains - We use the ISM appropriate dust grain model provided by the CLOUDY code \cite{10.1111/j.1365-2966.2004.07734,2006ApJ...645.1188W}. It includes both silicate and graphite components and has a size distribution and abundance that matches observed dust grain properties. Some metals from the gas phase are depleted as they are incorporated into the grain component, in particular calcium, aluminium, titanium, and iron.

\item  We use the UV/X-ray background model described in Haardt and Madau 2012 \cite{2012ApJ...746..125H}. This includes contributions from quasars, as well as the cosmic microwave background (CMB). 

\item Cosmic Ray Background - Cosmic rays provide a critical source of heating and ionization of the neutral gas. Furthermore, within the CLOUDY framework the inclusion of cosmic rays is critical to allow the code to maintain the chemistry network as the environment approaches lower temperatures at which the gas becomes molecular. The default background provided by the code is the model by Indriolo et al \cite{2007ApJ...671.1736I} with a mean ionization rate of $2 \times 10^{-16} \,~ \mathrm{s}^{-1}$; We note that the mean ionization rate indicates the fraction of atoms ionized per second in interstellar gas. Other authors (see \cite{2012ApJ...745...91I} and references within) however find more conservative rates as low as a $1-2 \times 10^{-18} \,~ \mathrm{s}^{-1}$, though more recent measurements seem to favour a value closer to $10^{-16} \,~ \mathrm{s}^{-1}$. In contrast McCall et al find a mean ionization rate of $1.2 \times 10^{-15} \,~ \mathrm{s}^{-1}$ along a galactic line of sight. We therefore consider a range of possible cosmic ray background ionization rates for our gas cloud models.

\item Density Profile - We assume a constant gas density for the cloud models presented here. While present surveys of these gas clouds reveal largely uniform gas cloud densities \cite{McClure-Griffiths:2013awa}, it would be interesting to extend our calculations for nonuniform gas cloud densities in future work. 
\end{itemize}

\begin{figure}
\centering
\includegraphics[scale=0.75]{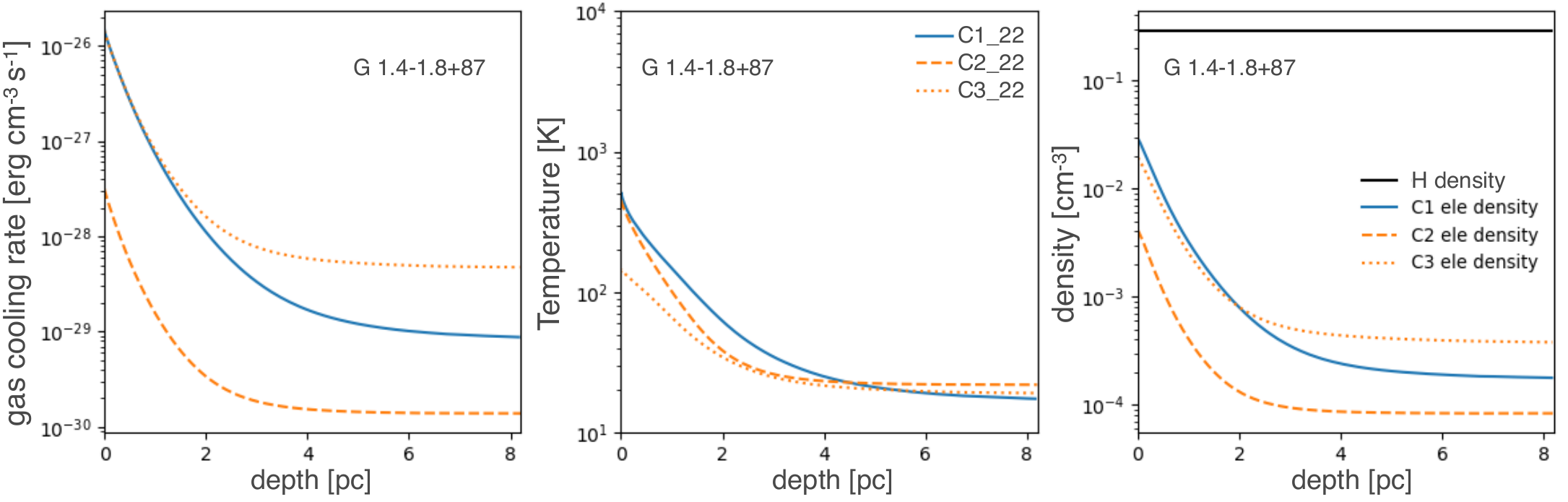}
\includegraphics[scale=0.75]{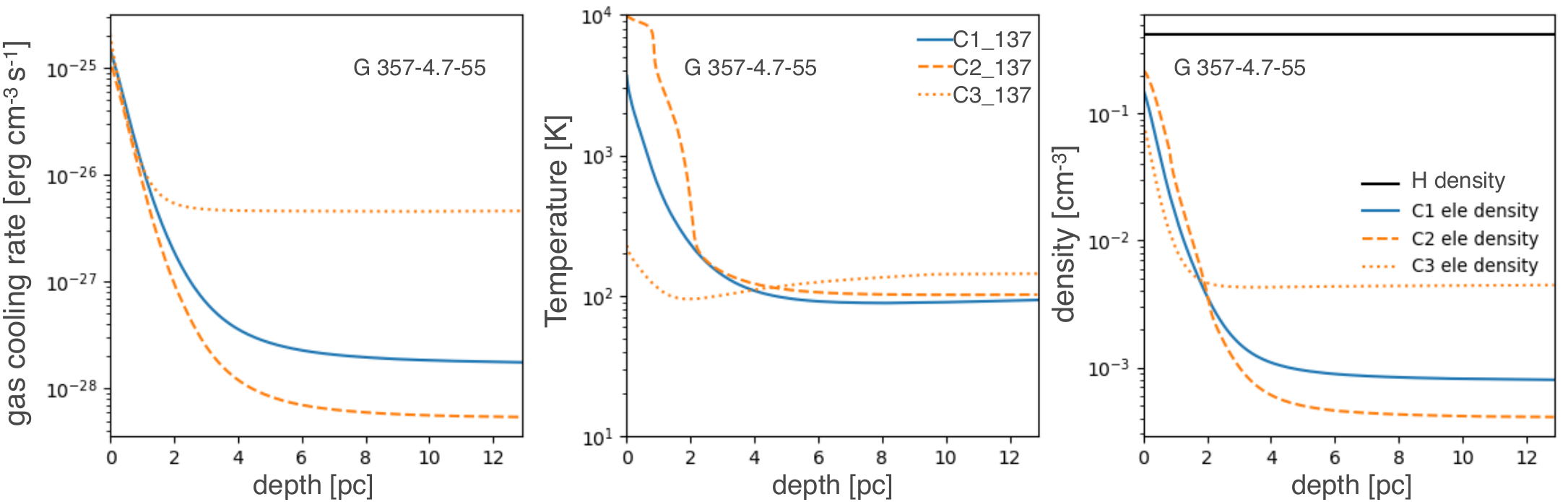}
\includegraphics[scale=0.75]{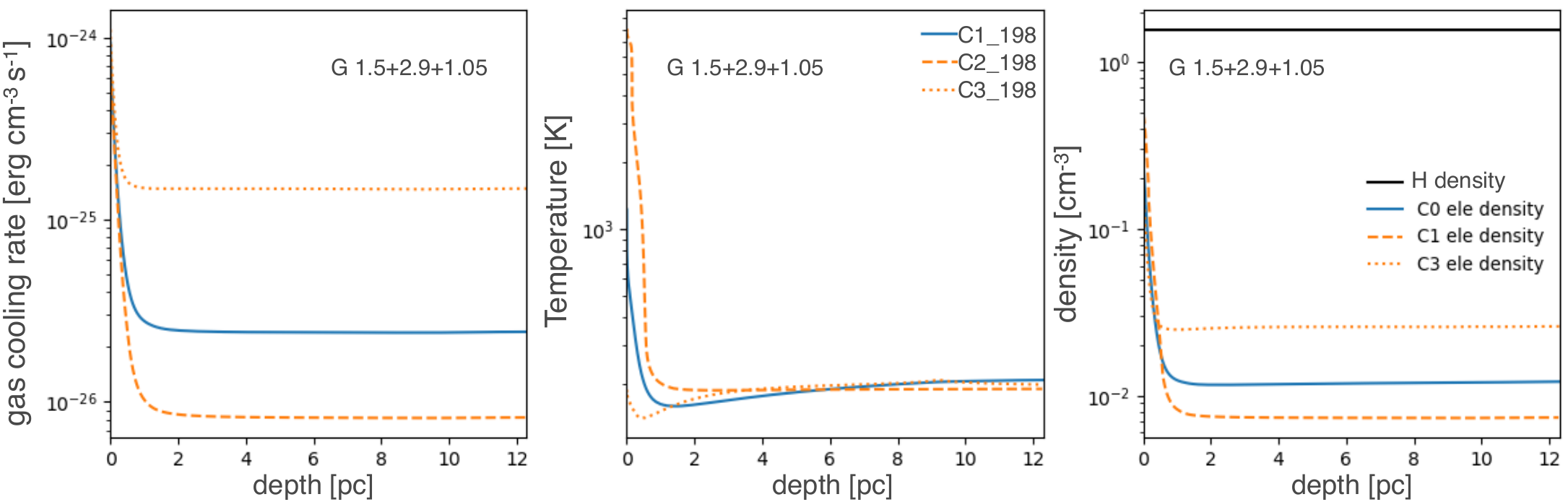}
\caption{CLOUDY models from top to bottom for clouds G1.4-1.8+87, G357.8-4.7-55 and G1.5+2.9+1.05 from \cite{McClure-Griffiths:2013awa}. Columns from left to right show the equilibrium cooling rate, temperature, and unbound electron (and hydrogen) number density for a given radial depth into the cloud, as measured from the surface. As detailed in Table 1, the radial depth of these clouds varies from 8-12 parsecs.  The solid blue, dashed and dotted orange lines correspond to the C1, C2, and C3 gas cloud models described in Table 1. In the density plots, the solid black line shows the constant hydrogen density.}
\label{fig:clouds}
\end{figure}

\subsection{Gas Cloud Models}
\label{subsec:outputs}
We here present CLOUDY outputs for the three coldest clouds found in the HI survey. Details of the clouds are given in Table 1. Columns detailing average temperature, density and radius are taken from \cite{McClure-Griffiths:2013awa}, while the metallicity, dust grain model, ultraviolet and cosmic ray backgrounds are settable parameters within CLOUDY. Since atomic transitions of ``metal" (heavier than helium) elements provide the dominant cooling mechanism, we explore the effect of varying gas cloud metallicity. For each cloud we consider one solar, one high and one low metallicity model, respectively called C1, C2 and C3, for a uniform gas density. The clouds were modeled as constant density spheres, although experimenting with oblate spheroid geometries did not substantially change the result. For each gas cloud, we tune the ultraviolet and cosmic ray parameters (which will physically tend to vary depending on the location of the cloud) to produce a cloud matching the observed average temperature.

\begin{table*}
\centering
\begin{tabular}{l  c c c c c c  c c c c}
\hline
DM       & $\bar{T}$ &  radius & $\bar{\rho}$ & $Z/Z_{\odot}$ & grains & UV & CR  & $\bar n_e$ & ave.~cooling \\
 Model   & [$\mathrm{K}$] &  [$\mathrm{pc}$]  & [$\mathrm{cm}^{-3}$] & &  &  & [$\mathrm{s}^{-1}$]  &  [$\mathrm{cm}^{-3}$]  & [$\mathrm{erg} \,\mathrm{cm}^{-3}\mathrm{s^{-1}}$]\\
\hline
C1-22 & $22$ &  $8.2$ & $0.29$ & $1$ & no & 0.1 & $1 \times 10^{-18} $ & $2.3 \times 10^{-4}$ & $1.9 \times 10^{-29}$\\
C2-22 & $22$  & $8.2$ & $0.29$& $0.1$ & no&$ 1.9 \times 10^{-3}$ &$ 1.9 \times 10^{-19} $&$ 9.7 \times 10^{-5}$& $1.6 \times 10^{-30}$\\
C3-22 & $22$ &  $8.2$ & $0.29$& $5$ & no & 0.1 & $ 5 \times 10^{-18}$ &$ 5.6 \times 10^{-4}$&$ 6.2 \times 10^{-28}$\\

\hline
C1-137 & $137$ & $12.9$& $0.421$ &  $1$ & yes & 1 & $ 5 \times 10^{-17}$ &$ 1 \times 10^{-3}$&$ 3.4 \times 10^{-28}$\\
C2-137 & $137$ &  $12.9$& $0.421$ & $0.1$ & yes & 1 & $ 3 \times 10^{-18}$& $ 5 \times 10^{-4}$ &$ 8.2 \times 10^{-29}$\\
C3-137& $137$  & $12.9$& $0.421$ & $5$  & yes & 1 &$ 1.9 \times 10^{-16}$& $ 6.2 \times 10^{-3}$& $ 6.1 \times 10^{-27}$\\

\hline
C1-198 & $198$ & $12.3$& $1.57$ & $1$ & yes & 1 & $ 2.9 \times 10^{-16}$ & $ 1.2 \times 10^{-2}$& $ 2.4 \times 10^{-26}$\\
C2-198 & $198$ & $12.3$& $1.57$  &  $0.1$ & yes & 1 & $ 1.1 \times 10^{-16}$ & $ 7.4 \times 10^{-3}$ & $ 8.2 \times 10^{-27}$\\
C3-198 & $198$ &  $12.3$& $1.57$  & $5$ & yes & 1 & $ 1.4 \times 10^{-15}$&$ 4.5 \times 10^{-2}$ &$ 1.5 \times 10^{-25}$\\

\hline 

\end{tabular}
\caption{Summary of the different gas cloud models simulated using CLOUDY, which match the properties of clouds G1.4-1.8+87, G357.8-4.7-55 and G1.5+2.9+1.05 from \cite{McClure-Griffiths:2013awa}. The average temperature ($\bar{T}$), gas cloud radius, and density ($\bar \rho$) are taken from McClure-Griffiths while the metallicity relative to solar metallicity $(Z/Z_\odot)$, presence of dust grains in the simulation, ultraviolet (UV) photon background flux relative to the standard normalization described in the text, the cosmic ray background ionization rate (CR), and the density profile parameter were parameters varied in the code. The average electron density ($\bar n_e$) and average cooling (ave.~cooling) rates are used for setting bounds on dark matter.}
\end{table*}

Two of the two key quantities required for setting dark matter bounds are the electron number density and the cooling rate. These are listed in the last two columns of Table 1 and presented in Figure 1. The three rows in Figure 1 correspond to the three gas clouds modeled, listed from coldest to hottest from top to bottom. Columsn from left to right show the equilibrium cooling rate (assuming only standard astrophysical sources), temperature, and electron densities for the gas cloud models previously mentioned. Note that the x axis shows the depth into the cloud from the illuminated surface rather than radius from the center of the object. In each plot, the blue solid, yellow dashed, and yellow dotted lines correspond to model C1, C2, and C3 (summarised in Table 1) respectively; these models assume different gas cloud metallicities as indicated. In the density plots, the solid black line indicated the uniform gas density. The other curves show the electron number density of the corresponding cloud models. 

As expected, models with higher metallicities provide more efficient cooling channels and therefore allow for a greater external energy input, whether it be from standard astrophysical sources or dark matter, to maintain the temperature observed in \cite{McClure-Griffiths:2013awa}. Higher metallicity clouds also result in higher electron number densities, because the metal species provide more readily ionized electrons to the gas. When setting bounds using these systems it is important to note that both the electron density and cooling rate are not derived independently, and there are some uncertainties in both the metallicity, dust grain and molecular content as well as the local UV radiation and cosmic ray background. Follow-up observations may allow these to be constrained further and therefore provide improved dark matter bounds. 

\subsection{Gas Cloud Bounds on Dark Matter}

In the following sections, we will present Galactic Center gas cloud bounds on dark matter. We note that these bounds are based on the fact that cold gas clouds cool predominantly via radiative cooling. As was pointed out in \cite{Bhoonah:2018wmw}, radiative cooling for gas clouds with temperatures $\sim 10-1000$ K is a monotonic decreasing function of temperature (see $e.g.$ Figure 4 in \cite{Maio:2007yf} and Figure 5 in \cite{DeRijcke:2013zha}). As a consequence of the monotonic decreasing form of gas cloud radiative cooling curves and the Second Law of Thermodynamics, there is a maximum possible heating rate for any gas cloud observed at fixed density and temperature ranging from $\sim 10-1000$ K. This implies the following bound on dark matter heating of cold gas clouds,
\begin{align}
VDHR \leq VCR
\end{align}
where the first term is the volumetric heating of the gas cloud by dark matter, and the second term is the volumetric cooling via radiative processes detailed earlier.

Because it is most likely that Galactic Center gas clouds have solar metallicity \cite{2013ApJ...777...19H}, especially given the relatively young age of the gas clouds in question ($\sim$ 10 Myr), we use model the solar metallicity model C1 for our gas clouds throughout the remainder of this paper to set bounds on dark sectors. As noted in \cite{Bhoonah:2018wmw}, models C2 and C3 yield similar bounds in the case of dark matter that predominantly interacts with electrons or iron, which is the case for dark matter considered in all three sections of this paper. This is because the dark matter bounds on interaction cross-sections scale linearly with both the gas cloud cooling rate and the electron or iron density. On the other hand, as can be verified from Table 1, the gas cloud cooling rate scales inversely with gas cloud metallicity (and by extension, iron density) and the gas cloud cooling rate also scales inversely with electron density. This makes the gas cloud heating bounds on dark matter interactions relatively insensitive to assumptions about gas cloud metallicity.

A few comments are in order concerning the dark matter density in the Galactic Center, as this does affect bounds on dark matter interactions with Galactic Center gas clouds. Technically, there is no direct evidence of dark matter in the central three kiloparsecs of the Milky Way, since this region is predominantly composed of baryonic matter \cite{Sofue:2013kja,Benito:2016kyp}. However, results from N-body simulations, hierarchical clustering, and dynamical halo structure considerations  \cite{Graham:2005xx}, indicate that the most plausible halos will have a shape similar to the eponymous NFW profile \cite{Navarro:1996gj}. In this study, we will use a generalized NFW profile as presented in \cite{Benito:2016kyp}, 
\begin{align}
\rho_{x}(r) = \rho_0 \left( \frac{r_0}{r} \right)^\gamma \left( \frac{r_s+ r_0}{r_s +r} \right)^{3-\gamma},
\end{align}
where $r_s = 20$ kpc is the standard scale radius of the Milky Way. We will use the generalized NFW paramaters in \cite{Benito:2016kyp} which were fit to match a morphological model of the stellar matter in the inner ``bulge" region of the Milky Way galaxy \cite{2009arXiv0903.0946V}. The parameters of model ``CjX" in \cite{Benito:2016kyp} are $r_0 = 8$ kpc, $\gamma \approx 1.03$, and $\rho_0 = 0.471~{\rm GeV/cm^3}$. This yields dark matter densities near Galactic Center gas clouds of approximately $\rho \sim 10~{\rm GeV/cm^3}$, which agrees well with standard halo profile parameters in the literature \cite{Bhoonah:2018wmw}. In the captions of bounds presented in this paper, we provide a simple prescription to re-scale bounds, for readers who wish to consider the effect of different background dark matter densities.

The line-of-sight distances of gas clouds G1.4-1.8+87, G357.8-4.7-55, and G1.5+2.9+1.05 from the Galactic Center are $R_{G1.4}=0.31$ kpc, $R_{G357}=0.75$ kpc, and $R_{G357}=0.41$ kpc respectively. Because the generalized NFW halo model predicts an increased dark matter density in the Galactic Center, when calculating Galactic Center gas cloud local dark matter densities, we will conservatively multiply these line-of-sight distances by a factor of $\sqrt{2}$ to account for their projected distance from the Galactic Center. Therefore, the projected distances we use for G1.4-1.8+87, G357.8-4.7-55, and G1.5+2.9+1.05 are $r_{G1.4}=0.44$ kpc, $r_{G357}=1.1$ kpc, and $r_{G357}=0.58$ kpc, respectively, implying dark matter densities near these three gas clouds of $\rho_{x,G1.4}=17 ~{\rm GeV/cm^3}$, $\rho_{x,G357}=6.6 ~{\rm GeV/cm^3}$, and $\rho_{x,G1.4}=13 ~{\rm GeV/cm^3}$, respectively. 

Finally, we note that throughout this document we will use a galactic center velocity dispersion of $\bar{v} \approx 180~{\rm km/s}$. This velocity dispersion is consistent with results in \cite{Sofue:2013kja}, and is on the low-end of velocity dispersion values allowed for by Milky Way dynamical considerations \cite{Benito:2016kyp}. Using this velocity dispersion will tend to produce conservative bounds in the case of dark matter-nucleon scattering and dark matter-electron scattering for heavy dark photon mediated dark matter (for which the dark matter induced gas cloud heating rate scales roughly as velocity cubed). On the other hand, this low velocity dispersion does produce slightly aggressive bounds in the case of very light dark photon mediated dark matter, considered at the beginning of section \ref{sec:subMevmed} (for which dark matter induced gas cloud heating scales inversely with velocity). In the latter case, we have verified that changing the velocity dispersion by a factor of two changes the bounds on the y-axis coupling parameters in Figure \ref{fig:light} by less than a factor of 1.2, which is not visible on the scale of the plot.  

\section{Ultra Light Dark Photon Dark Matter}
\label{sec:ldph}
Ultra light dark photon dark matter requires a rather simple extension of the Standard Model (SM), where the Standard Model gauge group is augmented by an extra local $U\left(1\right)$ symmetry. This model has, in addition to the Standard Model hypercharge, another abelian gauge boson, which we denote by $A^{\prime}$ and call the ``dark photon" \cite{Holdom:1985ag,Pospelov:2007mp,Pospelov:2008zw,Ackerman:mha,ArkaniHamed:2008qn,McDermott:2010pa,Davidson:2000hf}. The dark photon has a mass, and a kinetic mixing with the Standard Model hypercharge boson. For dynamical processes occurring in the sub-GeV range the physical $A^{\prime}$ field mixes predominantly with the Standard Model photon. The resulting Lagrangian is 
\begin{equation}
\mathcal{L} = \mathcal{L}_{SM}- \frac{1}{4}F_{\mu\nu}F^{\mu\nu} -\frac{1}{4}F^{\prime}_{\mu\nu}F^{\prime\mu\nu} + m^{2}A^{\prime}_{\mu}A^{\prime\mu} - \frac{e}{\left(1+\epsilon\right)^{2}}\left(A_{\mu} + \epsilon A^{\prime}_{\mu}\right)J_{EM}^{\mu},
\end{equation}  
Here, the kinetic part of the Lagrangian has been diagonalised, and we have adopted the convention of \cite{Dubovsky:2015cca} for the definition of the mixing parameter $\epsilon$. The interested reader is invited to consult \cite{Galison:1983pa} for a review of the form of Lagrangians with two local $U\left(1\right)$ gauge symmetries. The mass $m$ of the dark photon can be generated via the St\"uckelberg mechanism for simplicity, although it is straightforward to add an extra scalar field and generate $m$ via spontaneous symmetry breaking. The interaction part of this Lagrangian consists of the electromagnetic current $J_{EM}^{\mu}$ coupled to the photon and the dark photon, with the latter coupling suppressed by a factor of $\epsilon$ in the limit that $\epsilon \ll 1$. 

Much like axion dark matter, an ultra light dark photon is a plausible dark matter candidate, because it can provide a matter-like energy density via oscillations of the dark photon field. Ultra-light dark photon dark matter can be produced by cosmological excitation of its longitudinal or transverse field components as first considered in \cite{Nelson:2011sf,Arias:2012az,Graham:2015rva}. Assuming no additional couplings to lighter fields, ultra-light dark photon dark matter is meta-stable, since for $m_{A'} \ll 2m_e$, it decays to three photons with a rather long lifetime~\cite{Pospelov:2008jk} 
\begin{equation}
\tau_{A'} =  \frac{2^7 3^6 5^3 \pi^3}{17 \epsilon^2 \alpha_{\rm EM}^4 m_e} \left( \frac{m_e}{m_{A'}}\right)^9.
\end{equation}
It can be verified that even for an order one mixing, $\epsilon \sim O(1)$, so long as $m \ll  \text{keV}$, the dark photon is long lived enough to be a dark matter candidate. Recently, a number of studies have indicated that an excited scalar field can transfer energy to a light dark photon field, and that through this mechanism, even extremely light dark photons may constitute the entire dark matter abundance~\cite{Dror:2018pdh,Co:2018lka,Agrawal:2018vin}. 

In this work we place new bounds on ultra light dark photon dark matter for $m \leq 10^{-10}$ eV. We use the heating mechanism detailed in \cite{Dubovsky:2015cca}, which can be summarised as follows. An ultra light dark photon, through its mixing with the Standard Model photon produces an oscillating electric field which generates a current and dissipation in any medium that is not a perfect conductor. In our case cold gas clouds at the Galactic Center harbor unbound electrons and ions which will collide with each other after they are accelerated by the oscillating electric field. This altogether transforms dark photon potential energy into the kinetic energy of charged particles in cold gas clouds. We will only highlight the most important facets of the mechanism here -- the interested reader is encouraged to consult \cite{Dubovsky:2015cca} for a thorough derivation of the effect. 

We model the ionized part of our gas clouds as a non-relativistic plasma. For the purpose of setting bounds on dark photon dark matter, this is a conservative approximation, since we neglect collisions between the ionized and neutral component of our gas clouds, where these additional collisions would result in greater energy transferred from the dark photon field to the gas. For gas clouds reported in \cite{McClure-Griffiths:2013awa}, the plasma frequency, $i.e.$ the typical electrostatic oscillation frequency of electrons in response to a charge separation is
\begin{equation}
\omega_{p} = \sqrt{\frac{4\pi n_{e}}{m_{e}}} \approx 5 \times 10^{-13} \ eV ~\left(\frac{n_e}{2 \times 10^{-4}~{\rm cm^{-3}}} \right)^{1/2}
\end{equation}
where here we normalize to gas cloud G1.4-1.8+87's electron density \mbox{$n_{e} \approx 2\times10^{-4} ~{\rm cm^{-3}}$}, as given in Table 1.

Since we are considering a plasma medium and not empty space, there is a screening effect that limits the interaction range of electrons with any external electric field. This is the Debye length of the plasma (\mbox{$\lambda_{\rm d} = \sqrt{T_{\rm g}/(4 \pi \alpha_{\rm EM} n_{\rm e})} $}), which can be interpreted as the effective compton wavelength (related to the effective mass) of the dark photon in a plasma. 
If the dark photon mass is around the size of the plasma frequency, $m \sim \omega_{p}$ the dark photon can resonantly convert into ordinary photons. Because of this dark photon conversion process, cosmic microwave background data can be used to exclude dark photons with masses in excess of $\sim 10^{-14}$ eV \cite{Jaeckel:2010ni}. However, for masses below this, dark photons will not decay to photons in the early universe, and gas cloud heating by dark photon dark matter places leading bounds on this region of parameter space. 

\begin{figure}
\centering
  \includegraphics[width=1\textwidth]{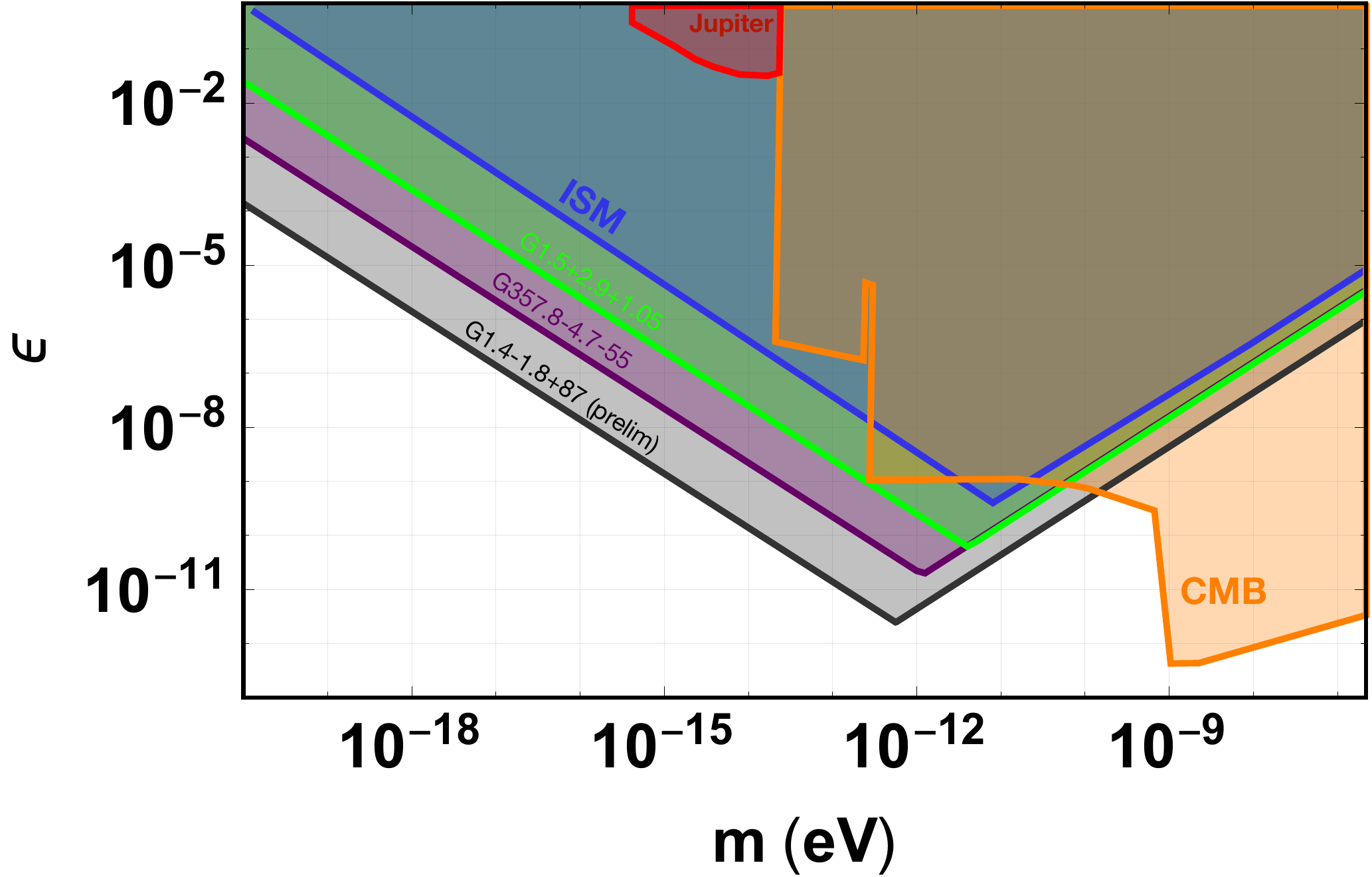}
 \caption{Bounds on ultra-light dark photons with mass $m$ and kinetic mixing parameter $\epsilon$, using three cold Galactic Center gas clouds, with parameters given in Table 1, and local dark matter densities according to a generalized NFW profile and projected distances from the Galactic Center detailed at the end of Section \ref{sec:gcprop}. Readers wishing to rescale these bounds for different dark matter background density models should note that the bound on $\epsilon$ scales as $\frac{1}{\rho_{x}^2}$. Bounds using gas cloud G1.4-1.8+87, using the temperature reported in \cite{McClure-Griffiths:2013awa} have been indicated as preliminary, see Section \ref{sec:add}. Different CMB limits from the decay of dark photons into Standard Model photons are also shown in orange \cite{Jaeckel:2010ni}. Bounds from satellite measurements of Jupiter's magnetic field are given in red \cite{Jaeckel:2010ni}. A constraint from heating of the Milky Way's interstellar medium by dark photons is shown in blue \cite{Dubovsky:2015cca}. The same mechanism, but applied to gas clouds with average temperature $137 \ K$ (G357.8-4.7-55 with an average cooling rate of $3.4 \times 10^-{28} {\rm  \ erg \ s^{-1} \ cm^{-3} }$) and $22 \ K$ (G1.4-1.8+87 with an average cooling rate of $1.9 \times 10^{-29}{\rm  \ erg \ s^{-1} \ cm^{-3} }$) are shown in purple and black respectively.}
  \label{fig:gcgcsi}
\end{figure}

To obtain the heating rate due to the ultra light dark photon field, the equations of motion of the two $U\left(1\right)$ vector fields are solved and combined with that of the non-relativistic plasma to obtain the frequency of the dark photon modes, $\omega$. The latter has a real and an imaginary part, $\omega = \omega_{h} + i\gamma_{h}$, and the imaginary part, $\gamma_{h}$ gives the volumetric heating rate Q, $i.e.$ the heating rate per unit volume
\begin{equation}
Q = 2 \vert \gamma_{h} \vert \rho_x
\end{equation}  
where $\rho_{x}$ is the dark matter density. In our case, the dark matter density is $\rho_x \sim 10 \ {\rm GeV \ cm^{-3}}$ for cold gas clouds near the Galactic Center \cite{Bhoonah:2018wmw}.  The asymptotic expressions for $\gamma_{h}$ in the limit that the dark photon mass is either smaller or larger than the gas cloud plasma frequency are
\begin{equation}\label{eq:heatrate}
\gamma_{h}  = \begin{cases} 
- \frac{\nu}{2}\left(\frac{m}{\omega_{p}}\right)^2\frac{\epsilon^{2}}{1+\epsilon^{2}}, & \textit{m $\ll \omega_{p}$} \\
- \frac{\nu}{2}\left(\frac{\omega_{p}}{m}\right)^{2}\frac{\epsilon^{2}}{1+\epsilon^{2}}, & \textit{m $\gg \omega_{p}$},
\end{cases}
\end{equation}
where $\nu$ is the collision frequency describing the interaction of electrons and ions in the plasma,
\begin{equation*}
\nu = \frac{4\sqrt{2\pi}\alpha_{EM}^{2}n_{e}}{3\sqrt{m_{e}T^{3}}}\log\left(\Lambda\right).
\end{equation*}
Here $\alpha_{EM}$ is the electromagnetic fine structure constant and $\Lambda$ is the Coulomb logarithm, 
$
\log\left(\Lambda\right) = \log\sqrt{\frac{4\pi T^{3}}{\alpha_{EM}^{3}n_{e}}}.
$
Even though the expressions for the heating rate in~\eqref{eq:heatrate} are asymptotic, we take them as hard cutoffs in establishing our bounds for light dark photons shown in~\eqref{fig:gcgcsi}. Further work would need to be done to accurately model dark photon heating of the plasma when $\omega_p \sim m$. We display new bounds on ultra light dark photon dark matter, obtained by equating the heating rate give in Eq. \eqref{eq:heatrate} to the cooling rates of gas clouds G357.8-4.7-55 and G1.4-1.8+87, as given in Table 1. 

\section{Sub-MeV Mediator Vector Portal Dark Matter}
\label{sec:subMevmed}

Besides being a dark matter candidate, dark photons can also serve as the mediator between dark matter and the SM. Indeed, this scenario has been extensively studied in the last decade~\cite{Feldman:2006wd,Feldman:2007wj,Pospelov:2007mp,Pospelov:2008zw,Essig:2013lka,Izaguirre:2015yja,Curtin:2014cca,Knapen:2017xzo}.  Cold Galactic Center gas clouds will prove especially sensitive to vector portal dark matter coupled through a sub-MeV mass dark photon, with intermediate strength couplings. Such dark matter evades detection by  terrestrial experiments, because it is moving too slowly to be detected after scattering with the Earth's atmosphere and crust, to excite electrons to detectable energies in existing experiments. 

We will consider a simple vector portal model to demonstrate that cold gas clouds can be used to explore dark matter models with light mediators. Our results indicate that other dark matter models coupled to the Standard Model through light mediators may also be constrained by cold Galactic Center gas clouds; this is left to future work. Here we take dark matter to be a Dirac fermion $\chi$ that communicates with the Standard Model via a kinetically mixed dark photon $A'$. Specifically, the Lagrangian we will be studying is the following:
\begin{equation}
\mathcal{L} = \mathcal{L}_{SM} - \frac{1}{2} m_{A'}^2 A'_{\mu} A^{'\mu}- \frac{1}{4} F'_{\mu \nu} F^{'\mu\nu} - \frac{\kappa}{2} F_{\mu\nu}F^{'\mu \nu} - g_D A'_\mu \bar \chi \gamma^\mu \chi
\end{equation} 
We note that in this section we will denote the kinetic mixing parameter as $\kappa$, in keeping with historical convention \cite{Pospelov:2007mp}, and so as not to confuse this with the $\epsilon$ mixing parameter in Section \ref{sec:ldph}; note that $\epsilon$ and $\kappa$ have different definitions. To take this gauge basis Lagrangian to the mass basis, we shift $A_\mu \to A_\mu - \kappa A'_\mu$, which results in an electromagentic current for $A'$ with a coupling proportional to $\kappa$.  The parameters in this simplified vector portal model are the dark photon mass $(m_{A'})$, the dark matter mass $(m_{\chi})$, the gauge coupling of the dark photon with dark matter $(g_D)$, and $\kappa$, which determines the coupling of dark photon with Standard Model fields. 

Since the interactions of the dark photon proceed through the Standard Model photon current, vector portal dark matter interacts with Standard Model particles charged under $U(1)_{\rm EM}$. For the coldest Galactic Center gas cloud of interest \mbox{G1.4-1.8+87}, we will only consider dark matter interactions with unbound electrons in the cloud; interactions with electrons bound to atoms and nuclei will be of subdominant importance, because the energy deposited on unbound electrons will be much larger. For the model detailed above, the cross section for dark matter scattering with unbound electrons has the following form
\begin{equation}
\sigma_{\chi e} = \frac{8 \pi \kappa^2  \alpha_D \alpha_{\text{EM}} \mu_{\chi e}^2}{(m_{A'}^2 + q^2)^2},
\end{equation}
with $\mu_{\chi e}  \equiv \frac{m_\chi m_e}{m_\chi+ m_e}$ being the reduced mass of the DM-electron system, $\alpha_D = g_{_D}^2/(4\pi)$, $q = \sqrt{2 E_{\rm nr} m_e}$ the momentum transfer between dark matter and electron, and $E_{\rm nr} \approx \mu_{\chi e}^2 v_{x}^2/m_e$ being the average energy transferred per elastic scattering interaction. Using cold gas clouds, a constraint on this model is obtained by requiring $VCR > VDHR \sim n_e n_{x} \sigma_{\chi e}v_{x} E_{\rm nr}$, where VCR is the cooling rate of the designated gas cloud per unit volume, and VDHR represents the DM heating rate per unit volume:
\begin{equation}
VCR >   \frac{ 8 \pi \kappa^2 \alpha_D \alpha_{\text{EM}} n_e n_{x}  \mu_{\chi e}^4}{m_e}  \int d^3v_{x } v_{x }^3   \frac{ B (v_{x},v_{\rm esc},y)} {\left(m_{A'}^2+ 2  \mu_{\chi e}^2 v_{x}^2\right)^2}. 
\label{eq:vcr1}
\end{equation}
The integral is taken from zero to the escape velocity of dark matter from the Milky Way, which is $\sim 0.002$ \cite{Piffl:2013mla}, with $y \equiv {\rm cos }~ \theta$ indicating the angle between the dark matter and the electron, and B is a Maxwellian DM velocity distribution defined in \cite{Bhoonah:2018wmw}, with normalization  $ \int d^3v_{x} B = 1$.  To gain some intuition for the bounds provided by Galactic Center gas clouds over a range of dark photon masses ($m_{A'}$), we will investigate two limits: 1) $m_{A'} \ll q$, and 2) $m_{A'} \gg q$, where in the first case the interaction of dark matter with electrons is through a long-range force and in the latter case it is a contact interaction. 

In the limit that $m_{A'} \to 0 $, we would expect an infrared divergence to arise in the integral in Equation \eqref{eq:vcr1}, since in the limit $v_x \rightarrow 0$ the integral diverges. This infrared divergence is regulated for dark matter scattering in cold gas clouds, because the dark photon will gain a thermal mass through its interactions with the gas cloud plasma as also discussed in Section \ref{sec:ldph}. The thermal mass can be found from the Debye length, which is \mbox{$\lambda_{\rm d} = \sqrt{T_{\rm g}/(4 \pi \alpha_{\rm EM} n_{\rm e})} $}. The Debye length indicates the scale at which the dark photon necessarily mediates finite range interactions in a cold gas cloud, compared to infinite range interactions it would mediate in empty space. 
Taking the thermal mass into consideration, we arrive at a well-defined infrared cutoff for the case of a very light $A'$. For $m_{A'} \ll q$
\begin{equation}
\kappa^2  \alpha_D \alpha_{\text{EM}} <  \frac{VCR}{n_e}  \left[ \frac{ 2 \pi n_{x}}{m_e}  \int d^3v_{x }  ~ \frac{  B (v_{x},v_{\rm esc},y)} {v_{x}} \log \left(\frac{2 \mu_{\chi e}^2 v_x^2}{(\text{max}[1/ \lambda_d,  m_{A'}])^2}\right)\right]^{-1}.
\label{eq:long-range}
\end{equation} 
The milli-charged DM is the special case of the scenario discussed here, where the vacuum mass of $A'$ is zero. If $m_{A'} > 1/\lambda_d$, then the infrared divergence is regulated by mass of dark photon. 

When the dark photon is larger than both the plasma frequency and the maximum momentum exchanged between the electron and dark matter, the dark matter electron scattering interaction becomes a contact ineraction. For $m_{A'} \gg q, 1/\lambda_d$,
\begin{equation}
\kappa^2  \alpha_D \alpha_{\text{EM}} <  \frac{VCR}{n_e } \left[ \frac{ 8 \pi  n_{x} \mu^4_{\chi e} }{m_e}  \int d^3v_{x}~  v_{x}^3 \frac{  B (v_{x},v_{\rm esc},y)} {m_{A'}^4} \right]^{-1},
\label{eq:contact}
\end{equation} 
which is well-defined in the sense that there are no divergences in this expression.

\begin{figure}[t!]
 \centering
  \includegraphics[width=.66\textwidth]{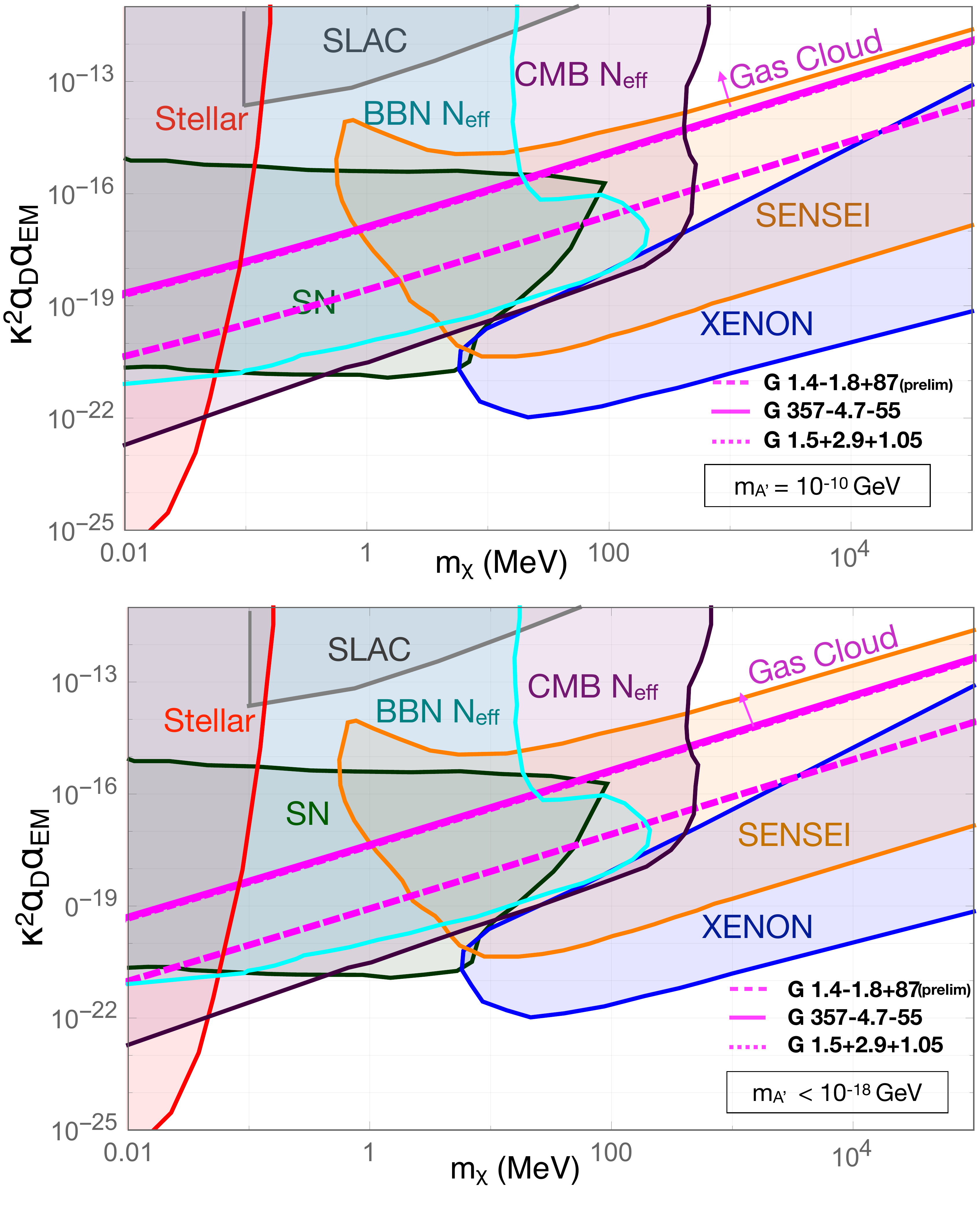}
  \caption{The constraint on the product of the couplings $ \kappa^2 \alpha_{_D} \alpha_{_\text{EM}}$ are shown in magenta (regions above each line are excluded), for vector portal dark matter heating of Galactic Center gas clouds with gas cloud parameters given in Table 1, and local dark matter densities according to a generalized NFW profile and projected distances from the Galactic Center detailed at the end of Section \ref{sec:gcprop}. Readers wishing to rescale these bounds for different dark matter background density models should note that the bound on $\kappa^2 \alpha_{_D} \alpha_{_\text{EM}}$ scales as $\frac{1}{\rho_{x}}$. Bounds using gas cloud G1.4-1.8+87, using the temperature reported in \cite{McClure-Griffiths:2013awa} have been indicated as preliminary, see Section \ref{sec:add}.  A few benchmark masses with $m_{A'} \ll \alpha m_e$ are shown. The constraints from the Galactic Center gas cloud apply up to coupling values given in Eqs.~\ref{eq:lightovb} and~\ref{eq:heavyovb}. The mass range $m_{A'} \ll \alpha m_e$ is chosen in this figure for ease of comparison with terrestrial experiments (SENSEI~\cite{Crisler:2018gci}, XENON~\cite{Essig:2017kqs}). Other constraints come from the SLAC millicharge search~\cite{Prinz:1998ua} shown in gray, supernova 1987A cooling~\cite{Chang:2018rso} presented in green, and the stellar cooling~\cite{Vogel:2013raa} shown in red. The constraints on effective relativistic degrees of freedom $N_{\rm eff}$ from BBN and CMB observations~\cite{Vogel:2013raa} are shown in cyan and purple, respectively.}
 \label{fig:light}
\end{figure}

\begin{figure}[t!]
 \centering
  \includegraphics[width=.66\textwidth]{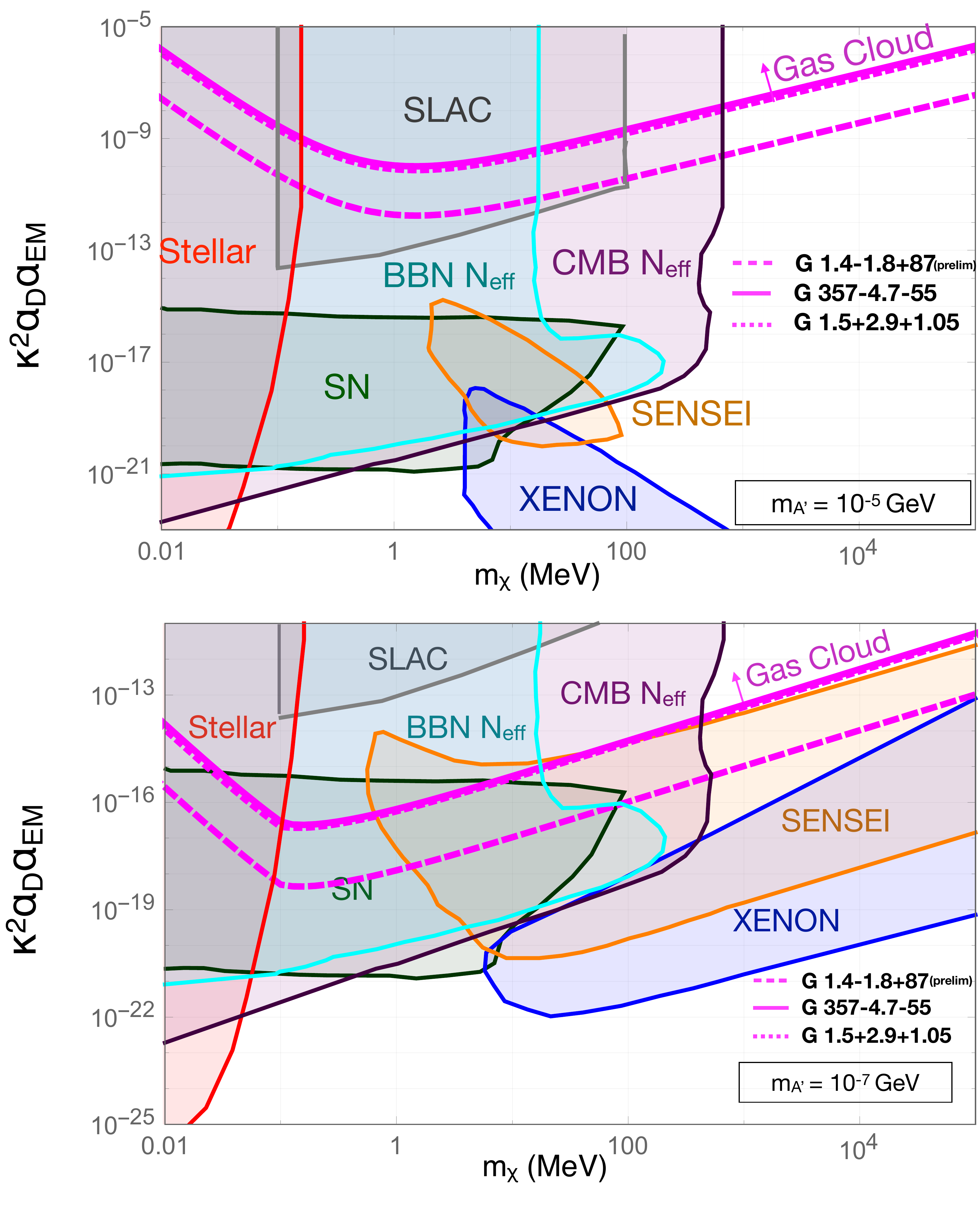}
  \caption{Constraints on the product of the couplings $ \kappa^2 \alpha_{_D} \alpha_{_\text{EM}}$ are shown in magenta (regions above each line are excluded), for vector portal dark matter heating of gas clouds with cloud parameters given in Table 1, and local dark matter densities according to a generalized NFW profile and projected distances from the Galactic Center detailed at the end of Section \ref{sec:gcprop}. Readers wishing to rescale these bounds for different dark matter background density models should note that the bound on $\kappa^2 \alpha_{_D} \alpha_{_\text{EM}}$ scales as $\frac{1}{\rho_{x}}$. Bounds using gas cloud G1.4-1.8+87, using the temperature reported in \cite{McClure-Griffiths:2013awa} have been indicated as preliminary, see Section \ref{sec:add}.  A few benchmark masses with $m_{A'} > \alpha m_e$ are shown. We note that constraints from Galactic Center gas cloud are effective up to coupling values given in Eq.~\ref{eq:heavyovb}. Terrestrial direct detection bounds \cite{Crisler:2018gci,Essig:2017kqs} are comparatively weak in this regime. We show the bounds for the two cases of $m_{A'} = 10^{-7} \ \text{GeV}$ and $m_{A'} =10^{-5} \ \text{GeV}$. For an explanation of the feature apparent at $m_\chi = 10^{-4} {\rm GeV}$ for $m_{A'} = 10^{-7} {\rm ~GeV}$, see the text. Other bounds (SLAC~\cite{Prinz:1998ua},  Supernova ~\cite{Chang:2018rso}, $N_{\rm eff}$ from BBN and CMB~\cite{Vogel:2013raa}, and stellar constraints~\cite{Vogel:2013raa}) are indicated.  }
 \label{fig:heavy}
\end{figure}

In Figures \ref{fig:light} and \ref{fig:heavy}, we plot galactic gas cloud constraints on vector portal models, using Eq.~\eqref{eq:vcr1}, gas cloud parameters given in Table 1, particularly the average volumetric cooling rate obtained for gas cloud G1.4-1.8+87, $VCR = 1.9 \times 10^{-29}~{\rm ergs ~cm^{-3}~s^{-1}}$. In Figure \ref{fig:light} we note that for the mass choice $m_{A'} = 10^{-7}~{\rm GeV}$, the bound shifts dramatically at $m_\chi = 10^{-4} ~{\rm GeV}$. This is because at this mass value, the dark photon mass is roughly equal to the momentum exchanged between the dark matter and electron, $m_{A'} \sim q \sim m_\chi v_x$. Put another way, at this dark matter mass, the dark matter-electron scattering dynamics shift from long-range to contact interactions; therefore the relevant bound shifts from Eq.~\eqref{eq:long-range} to Eq.~\eqref{eq:contact}.

In Figures \ref{fig:light} and \ref{fig:heavy} and the preceding treatment, we have considered relatively small couplings between dark matter and electrons. However, if the coupling of dark matter to electrons ($\kappa^2  \alpha_D \alpha_{\text{EM}}$) is sufficiently large, bounds from Galactic Center gas clouds will no longer apply.  Dark matter with a large enough coupling to electrons will deposit most of its kinetic energy into electrons near the surface of the gas cloud, and may not appreciably heat the interior region. To estimate this effect, we will require that dark matter retain at least half of its initial kinetic energy by the time it reaches the center of the gas cloud. The energy transfer of a dark matter to the cloud is simply $dE/dr = n_e \sigma_{\chi e} E_{\rm nr}$. Given that the depth of the gas cloud G1.4-1.8+87 is $ r= 8$ pc, the coupling for which dark matter loses half its kinetic energy after traveling the radius of the cloud is
\begin{equation}
\kappa^2  \alpha_D \alpha_{\text{EM}} \sim 10^{-7} \left(\frac{m_{\chi}}{\text{GeV}}\right),
\label{eq:lightovb}
\end{equation}
for a long range interaction $(m_{A'} \ll q)$, and 
\begin{equation}
\kappa^2  \alpha_D \alpha_{\text{EM}} \sim  1 \times \left(\frac{m_\chi}{\text{GeV}}\right) \left(\frac{0.5~ \text{MeV}}{ \mu_{\chi e}}\right)^4 \left( \frac{m_{A'}}{\text{10~keV}}\right)^4,
\label{eq:heavyovb}
\end{equation}
for an interaction with $m_{A'} \gg q$. 

Terrestrial direct detection experiments are also only sensitive to dark photon mediated dark matter with sufficiently weak interactions, so that dark matter passes through the Earth's atmosphere and crust without losing too much of its kinetic energy. As a consequence, such terrestrial experiments are sensitive to small values of the dark-visible photon mixing parameter and dark sector gauge coupling constant. Fig.~\ref{fig:light} shows the comparison of gas cloud bounds with other astrophysical and terrestrial experiments. One significant astrophysical bound is obtained from Supernova1987A~\cite{Chang:2018rso} (green region in Fig.~\ref{fig:light}), since dark photon mediated dark matter can be produced during the implosion of the nascent neutron star. If the dark matter had left the supernova in appreciable quantities, the supernova would have cooled faster than was observed. For Supernova 1987A, any extra sources of energy loss per unit mass has to be smaller than $ 10^{19} \text{erg/g/s}$ at plasma temperatures of $\sim 10~{\rm MeV}$ \cite{Raffelt:1996wa}. If the dark matter is too strongly interacting, it will be trapped inside the supernova and cannot contribute to its cooling. Therefore, SN1987A also has an upper limit for the constraint it provides, evident in Figures \ref{fig:light} and \ref{fig:heavy}. This bound can be sensitive to the effective mass of dark photon. However, for dark photon masses $m_{A'} < T_{\rm SN} \sim 20 \ \text{MeV}$, the thermal mass of dark photon in the hot supernova plasma dominates, and the bounds do not change by varying bare $A'$ mass.  

Another astrophysical observation that constrains the parameter space comes from red-giant helium burning and white dwarf stars based on stellar energy loss~\cite{Brummer:2009cs, Davidson:1993sj,Davidson:2000hf,Vogel:2013raa}. Dark matter particles can be produced though an exchange of a dark photon in the stellar interior and leave the star, resulting in a faster cooling rate for these stars. The bounds coming from stellar observations are shown in red in Fig.~\ref{fig:light}. The effective number of relativistic degrees of freedom during big bang nucleosynthesis (BBN) and cosmic microwave background also provide substantial bounds on the parameter space~\cite{Vogel:2013raa}. 

Another important bound comes from the electron beam dump experiment at SLAC~\cite{Prinz:1998ua}. This experiment consisted of a $20 \ \text{GeV}$ electron-beam impinging upon a set of fixed aluminum plates. A pair of dark matter particles can be produced via the exchange of an off-shell $A'$: $e^- N \to  e^- N A^{'*} \to e^- N \bar \chi \chi$. The dark matter would then traverse through a $179 $ m wide hill, followed by $204-m$ of air, before scattering off of electrons detected in an electromagnetic calorimeter. We might also consider bounds from production of the dark photon itself. However, if the dark photon is produced on-shell and remains stable until reaching the detector, it will leave no detectable signature in this experiment. So long as $ m_{A'} < 2m_e$ and $m_{A'} < 2m_{_\chi}$, the bounds from the SLAC experiment on our vector portal dark matter model will only arise from production of dark matter.

Terrestrial direct detection experiments traditionally provide strong bounds on dark matter heavier than a GeV. However, while XENON10 was intended to constrain dark matter-nucleon scattering, it showed sensitivity to single ionized electrons~\cite{Essig:2017kqs}, and could thereby bound dark matter-electron interactions for dark matter masses as small as a few MeV. New proposals, using XENON10 as a proof-of-principle, have suggested the usage of semi-conductors due to their lower band gap. These will probe even lighter dark matter, and also have an enhancement in the event rate for heavier dark matter masses. One such experiment is SENSEI~\cite{Crisler:2018gci}, whose results are shown in Figure \ref{fig:heavy}.

At electron ionization experiments like SENSEI and XENON10, the relevant electrons are often bound to atoms. The typical velocity of a bound electron is $v_e \sim \alpha_{EM}$. Thus there is often a minimum threshold momentum exchange to dislodge the electron and detect dark matter-electron scattering: $q \sim \alpha_{EM} m_e$. As a result, these electron ionization experiments have taken to quoting electron scattering cross sections with the form
\begin{equation}
\bar \sigma_e =   \frac{ 8 \pi  \kappa^2 \alpha_D  \alpha_{\rm EM} \mu_{\chi e}^2}{(m_{A'}^2 + q^2)^2}=\frac{ 8 \pi  \kappa^2 \alpha_D  \alpha_{\rm EM} \mu_{\chi e}^2}{(m_{A'}^2 + (\alpha_{EM} m_e)^2)^2} ,
\label{eq:exp}
\end{equation}
since smaller momentum exchanges than $q = m_e \alpha_{EM}$ would not dislodge an electron. To translate the bounds provided by these experiments, which are given in terms of the above cross section, we must also multiply $\bar \sigma_e$ by a form factor $|F(q)|^2$, which contains the momentum transfer dependence of the interaction. For example, for a point-like interaction $m_{A'} \gg \alpha m_e$, the momentum transfer can be neglected, and $F(q) =1$. For interactions with an ultra light mediator, on the other hand, the form factor is $ F(q) = (\alpha m_e/q)^2$. In Figure~\ref{fig:light}, we have only considered $ m_{A'} \ll \alpha m_e \sim 10^{-6} \ \text{GeV}$.

The comparison of bounds for heavier dark photons ($m_{A'} \gtrsim \alpha_{EM} m_e$) is presented in Fig.~\ref{fig:heavy}. As is evident, there is some parameter space where cold gas clouds provide the prevailing bound. Gas cloud bounds, which are presented with magenta lines, are particularly important for $m_{_\chi} \gtrsim 1 \ \text{GeV}$. The astrophysical constraints as well as the bounds from SLAC milli-charged experiment and SN1987A remain the same. Finally, we note that there are likely to be bounds (derived at some point in the future) from the non-observation of spectral distortions of the CMB power spectrum, which complement cold Galactic Center gas cloud bounds on dark matter-electron scattering via a light vector portal mediator \cite{Dvorkin:2013cea,Gluscevic:2017ywp,Xu:2018efh}.

\section{Strongly Interacting and Composite Dark Matter}
\label{sec:strong}

There are a number of models which predict a large cross section for dark matter scattering with nuclei. For example, dark matter charged under the Standard Model SU(3) gauge group, $i.e.$ color-charged dark matter \cite{Raby:1997pb,Kang:2008ea,DeLuca:2018mzn}, near-Planck mass dark matter~\cite{Davoudiasl:2018wxz}, monopole dark matter ($e.g.$ \cite{Khoze:2014woa}), and dark matter composed of many constituent states, also known as composite dark matter \cite{Frieman:1988ut,Kusenko:1997si,Krnjaic:2014xza,Detmold:2014qqa,Hardy:2015boa,Grabowska:2018lnd,Coskuner:2018are}.

One substantial advantage cold gas clouds have in searching for heavy dark matter is a large admitted flux of dark matter. The  gas clouds observed at the Milky Way Galactic Center reached thermal equilibrium over the course of millions of years, and have radii $r_g \sim 10~{\rm pc}$. This means that, requiring a flux ($N_f$) of at least ten dark matter objects of mass $m_x$ pass through the cloud over a million years ($t_g$), Galactic Center gas clouds are sensitive to dark matter masses up to $m_x \sim \pi r_g^2 \rho_x v_x t_g /N_f$,
\begin{align}
m_x \simeq  3 \times 10^{60} ~{\rm GeV}~\left( \frac{r_g}{10~{\rm pc}} \right)^2 \left( \frac{\rho_{x}}{10~{\rm GeV/cm^3}} \right) \left( \frac{v}{0.001 c} \right)\left( \frac{t_g}{10^6~{\rm yrs}} \right)\left( \frac{10}{N_f} \right) .
\end{align}
For objects so massive -- one clear candidate would be primordial black holes -- heating by gravitational processes like dynamical friction should be considered. In this study, we restrict our attention to dark matter masses up to $\sim 10^{32}$ GeV, and assume that the dark matter primarily couples to baryons via a non-gravitational interaction.

\begin{figure}
\includegraphics[scale=1.4]{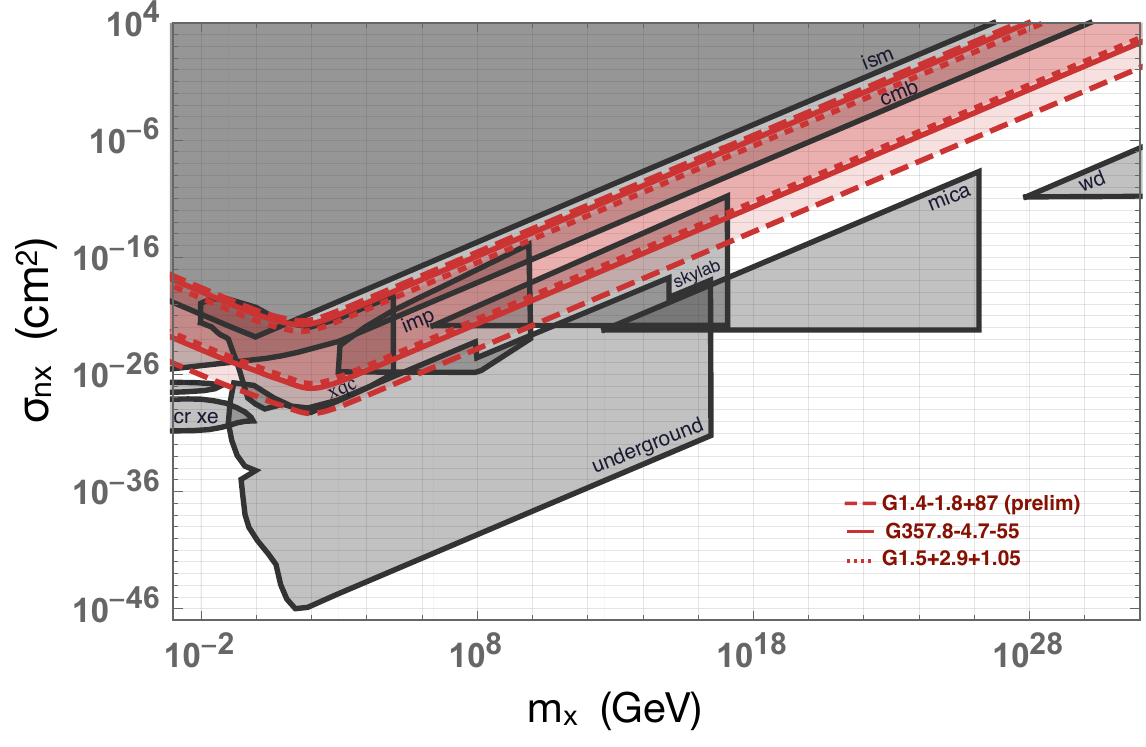}
\caption{Existing bounds on the spin-independent dark matter-nucleon scattering cross section are shown, along with bounds derived from cold gas clouds at the Galactic Center, for dark matter masses ranging from MeV to $10^{32}$ GeV, and local dark matter densities according to a generalized NFW profile and projected distances from the Galactic Center detailed at the end of Section \ref{sec:gcprop}. Readers wishing to rescale these bounds for different dark matter background density models should note that the bound on $\sigma_{nx}$ scales as $\frac{1}{\rho_{x}}$. Bounds using gas cloud G1.4-1.8+87, using the temperature reported in \cite{McClure-Griffiths:2013awa} have been indicated as preliminary, see Section \ref{sec:add}.  Note that cold gas clouds bounds can be accurately extrapolated out to $\sim 10^{60}$ GeV; we have truncated the plot for the sake of clarity. From left to right, bottom to top, prior bounds on dark matter-baryon scattering shown in gray are as follows: Cosmic ray accelerated dark matter is excluded by searches at Xenon1T (cr xe) and MiniBoone (cr mini) \cite{Bringmann:2018cvk}. Spectral distortions of the cosmic microwave background exclude stronger dark matter-baryon couplings (cmb) \cite{Gluscevic:2017ywp,Dvorkin:2013cea}. Interstellar gas cooling constraints (ism) were first derived in \cite{Chivukula:1989cc}. The lower bounds from underground direct detection experiments and particularly XENON1T \cite{Aprile:2017iyp} (underground) is combined with overburden upper bounds which were recently derived and summarized in \cite{Kavanagh:2017cru}. The X-ray Calorimetry Rocket did not observe dark matter events in its calorimeter (xqc) over its ten minute flight  \cite{Erickcek:2007jv}. Results from an as-yet unpublished analysis \cite{Wandelt:2000ad} of Interplanetary Monitoring Platform data \cite{imp8} are displayed (imp). A charged cosmic ray search using skylab's plastic etch detectors was recast for a dark matter bound in \cite{Starkman:1990nj}; this analysis (skylab) was later amended in \cite{McGuire:1994pq}. A bound from the longevity of white dwarf stars in the Milky Way is also shown (wd) \cite{Graham:2018efk}. Non-observation of tracks in ancient mica excludes dark matter with mass $10^{12}-10^{26}$ GeV, \cite{Bramante:2018tos,Price:1986ky,Jacobs:2014yca}.}
\label{fig:totsi}
\end{figure}

The sensitivity of underground experiments \cite{Aprile:2017iyp,Cui:2017nnn,Akerib:2016vxi,Amaudruz:2017ekt,Amole:2017dex,Arnaud:2017bjh,Agnese:2014aze,Bramante:2018qbc,Bramante:2018tos}, above-ground searches \cite{Rich:1987st,McGuire:1994pq,Bernabei:1999ui,Wandelt:2000ad,Albuquerque:2003ei,Zaharijas:2004jv,Mack:2007xj,Erickcek:2007jv,Kouvaris:2014lpa,Davis:2017noy,Dick:2017mgd,Mahdawi:2017cxz,Kavanagh:2017cru,Hooper:2018bfw,Emken:2018run}, and cosmological surveys \cite{Dubovsky:2003yn,McDermott:2010pa,Dvorkin:2013cea,Vogel:2013raa,Gluscevic:2017ywp,Xu:2018efh,Slatyer:2018aqg} to dark matter's interactions with baryons are summarized in Figure \ref{fig:totsi}. Bounds are expressed in terms of the per-nucleon scattering cross section, for ease of comparison between experiments, since target nuclei at these experiments vary. For example, argon, fluorine, and xenon are typical nuclear targets at underground direct detection experiments, while oxygen is a predominant nuclear target in ancient mica and Skylab's plastic etch detectors. Our treatment here assumes that dark matter couples equally to protons and neutrons in nuclei. Then in the case that dark matter couples directly to nucleons through a ``contact" interaction, $e.g.$ the mass of the boson mediating the interaction is much greater than the momentum exchange, the spin-independent dark matter-nucleon interaction is given in terms of the nuclear interaction as \cite{Lewin:1995rx}
\begin{align}
\sigma_{nx} = \left(\frac{\mu_{nx}}{\mu_{Nx}~A} \right)^2 ~\sigma_{N x} \frac{1}{F_{A}^2(E_{nr})}
\end{align}
where $\mu_{nx} \equiv \frac{m_x m_n}{m_x + m_n},\mu_{Nx} \equiv \frac{m_x m_N}{m_x + m_N}$ are reduced masses, $m_n$ is the nucleon mass, $m_N$ is the nuclear mass, $m_x$ is the dark matter mass, $A$ is the number of nucleons in the nucleus, and $\sigma_{Nx}$ is the dark matter-nuclear scattering cross section. The dark matter - nuclear scattering form factor for a lab frame nuclear recoil energy of $E_{nr}$ is given by
\begin{align}
F_A^2(E_{nr}) = \left(\frac{3 J_1(qr)}{qr} \right)^2 e^{-s^2q^2},
\end{align}
where the size of the nucleus is approximately $ r=\sqrt{r_n^2-5s^2}$ for nuclear skin depth $s = 1~{\rm fm}$ and nuclear radius $r_n = 1.2~{A^{1/3}}~{\rm fm}$, $J_1$ is the first Bessel function, and the momentum transfer in the scattering interaction is $q = \sqrt{2 m_N E_{nr}}$.

Bounds on dark matter interactions with baryons from gas clouds are obtained by requiring that the volumetric rate of dark matter heating via scattering with baryons, is less than volumetric cooling rate of the coldest gas cloud observed in the Galactic Center, $VDHR < VCR$ for gas cloud model $C1_22$ in Table 1. Specifically, the volumetric heating rate is given by $VDHR = n_x n_A \sigma_{Nx} v_x E_{nr}$, where $n_x = \rho_x/m_x$ is the number density of dark matter and $n_A$ is the number density of an atomic element in the gas cloud. Summing over contributions from scattering with hydrogen, helium, oxygen, and iron in the gas cloud, with relative mass fractions given assumed to be the solar mass abundances of $\{f_H,f_{He},f_O,f_C,f_{Fe}  \} = \{0.71,0.27,0.01,0.004,0.0014  \}$, this implies the following bound on the per-nucleon scattering cross section
\begin{align}
\sigma_{nx} < \frac{VCR}{n_n} \left( \sum_A \frac{  f_A \mu_{Nx} A^2n_x m_n}{\mu_{nx} m_N^2} \int d^3v_x~ v_x^3 F_A^2(E_{nr})B(v_x,v_{esc},y) \right)^{-1},
\end{align}
where $B(v_x,v_{esc},y)$ is the Maxwell-Boltzmann distribution given in \cite{Bhoonah:2018wmw}. In figure \ref{fig:totsi}, we find that cold Galactic Center gas clouds provide new bounds on baryonic dark matter interactions, particularly for dark matter masses in excess of $10^9$ GeV.

In order to effectively heat the entire gas cloud, dark matter must be able to move with a roughly constant velocity through the gas cloud. For strong enough dark matter interactions, the dark matter will have depleted its kinetic energy so much before reaching the gas cloud interior, that it will be incapable of appreciably heating the majority of the gas cloud. Indeed, as is evident in Section \ref{sec:gcprop} and Figure \ref{fig:clouds}, galactic gas clouds will have relatively hot exteriors. Therefore, if dark matter is substantially slowed in the outer layers of the gas cloud, it would not heat the central region of the gas cloud effectively. Future work may wish to model this explicitly, and find how the average and radially-distributed gas cloud temperature is affected by dark matter which is substantially slowed as it travels through a gas cloud. In our treatment here, to account for this ``overburden" effect in our Galactic Center gas cloud treatment, we will require that the dark matter retain at least half its kinetic energy after it has traveled to the center of the gas cloud. The formula describing the average depletion of dark matter kinetic energy as it travels through the gas cloud is given by ($e.g.$ \cite{Bramante:2017xlb})
\begin{align}
\frac{E_f}{E_i} = \prod_A \left(1- \frac{4z m_N m_x}{(m_N + m_x)^2} \right)^{n_A \sigma_{Nx} r},
\end{align}
where $\frac{E_f}{E_i}$ is the ratio of the final to initial kinetic energy after scattering with the gas cloud, $n_A \equiv f_A n_n \frac{m_n}{m_A}$ is the number density of nucleus $A$, $r$ is the distance the dark matter travels in the gas cloud, and $z \in (0,1)$ is a kinematic factor encapsulating the scattering angle: it ranges from zero to one for glancing to head-on collisions. The term $\frac{4z m_N m_x}{(m_N + m_x)^2}$ is the fraction of the dark matter's kinetic energy depleted by each scatter. In computing the critical cross section at which dark matter loses half its kinetic energy before reaching the center of the cloud, we take $z=0.5$. The solar metallicity mass fractions given above ($f_A$), along with the radial depth $r=8~{\rm pc}$ and hydrogen density $n_n \approx 0.3 ~{\rm cm^{-3}}$ of gas cloud G1.4-1.8+87 were used to obtain an upper bound on the per-nucleon scattering cross section constrained by Galactic Center gas clouds, shown in Figure \ref{fig:totsi}.

\section{Conclusions}
\label{sec:disc}
We have used Galactic Center gas clouds to place novel bounds on a number of dark matter scenarios, including ultra-light dark photon dark matter, sub-MeV mediated vector portal dark matter, and dark matter that scatters elastically with baryons. After presenting the first detailed study of Galactic Center gas clouds' composition, thermal properties, cooling rates, electron density, metal density, and ionization fractions, we used the derived cooling rates and ionization fraction to set bounds. Remarkably, we have found that Galactic Center gas clouds are excellent detectors for dark matter models spanning the entire range of plausible dark matter masses. Ultra-light dark photon dark matter ($\sim 10^{-22}$ eV), vector portal dark matter ($\sim {\rm keV-TeV}$ mass), and super heavy ($\lesssim 10^{60}$ GeV) dark matter that interacts with baryons all lie at the frontier of cold gas cloud detection.

Besides establishing cold gas clouds as excellent dark matter detectors, we have also begun the process of developing accurate models for the cold gas clouds recently discovered at the Galactic Center. The interstellar gas code CLOUDY has been adapted to model cold gas clouds matching the temperature, density, and size of the coldest clouds found near the center of the Milky Way galaxy. Besides validating the estimates of gas cloud cooling used in prior work (\cite{Bhoonah:2018wmw}), we explicitly determined the effect of super-solar and sub-solar metallicities on their cooling rates, ionization, and the intra-cloud distribution of each of these quantities. Cold gas cloud modelling will be useful for validating any future cold gas cloud \emph{detections} of dark matter. In particular, dark matter will alter the thermal properties of gas cloud interiors. In future work, it will be interesting to consider how cold Galactic Center gas clouds' internal thermal structure can be used as a diagnostic of dark matter interactions.

\section{Addendum}
\label{sec:add}
After this paper was completed, some work appeared commenting on Galactic Center gas clouds' suitability for setting bounds on dark matter interactions. Regarding gas cloud G1.4-1.8+87, which was reported to have a temperature $T \leq 22~{\rm K}$ in \cite{McClure-Griffiths:2013awa,2018ApJ...855...33D}, the authors of \cite{Farrar:2019qrv} have questioned whether G1.4-1.8+87 might not have the cold, $T \leq 22~{\rm K}$ gas core reported in \cite{McClure-Griffiths:2013awa,2018ApJ...855...33D}. A number of follow-on observations and studies are planned. While the analysis in \cite{McClureGriffiths:2012et,McClure-Griffiths:2013awa} indicates G1.4-1.8+87 has a $T \leq 22~{\rm K}$ core with the statistical significance required to satisfy data quality cuts. We have nevertheless indicated all bounds in this document that use G1.4-1.8+87 as ``preliminary," pending further investigation as there are concerns that the 22 K temperature is due to a single-channel fluctuation.

Separately, the authors of \cite{Wadekar:2019xnf} have inquired about the effect of a $\sim 200 ~{\rm km/s}$, $T \sim 10^6~{\rm K}$ Galactic Center wind, hypothesized and modeled in \cite{McClure-Griffiths:2013awa} to explain the distribution of Galactic Center gas clouds near the inner kpc of the Milky Way. Specifically, the authors of \cite{Wadekar:2019xnf} have questioned whether Galactic Center gas clouds are in heating/cooling ``equilibrium," given the hypothetical presence of a hot Galactic Center gas wind. 

It is important to note that while a new gas cloud analysis recently presented in \cite{Wadekar:2019xnf} {\em does} assume gas cloud heating/cooling equilibrium when setting bounds, the bounds presented in this paper and in our prior work \cite{Bhoonah:2018wmw} do not assume heating/cooling equilibrium of gas clouds. Rather, as explained in Section \ref{sec:gcprop} and Ref.~\cite{Bhoonah:2018wmw}, our bounds rely on gas cloud cooling being a monotonic decreasing function of temperature, for gas clouds with fixed density and temperatures $\lesssim 1000$ K. The basic logic is that, observing such a cold gas cloud, regardless of whether it is presently heating or cooling overall, implies a maximum possible historic heating rate for that gas cloud, or else it would not have been capable of cooling to such a low temperature.

This means that, insofar as the density of the gas clouds is approximately constant, one need not assume heating/cooling equilibrium to trust the bounds presented in this paper. Nevertheless, we find that standard analysis of Galactic Center gas clouds and the hypothetical surrounding wind, indicates that the Galactic Center gas clouds used in this study are not disrupted by the wind hypothesized in \cite{McClure-Griffiths:2013awa}. Indeed, this is the explicit conclusion drawn in \cite{McClure-Griffiths:2013awa}. One test for whether gas clouds are disrupted by a surrounding wind is to compare the shock cooling time, which is the time for the cloud to radiate an $\mathcal{O}(1)$ fraction of its kinetic energy in a region shocked by a hot wind, with the cloud crushing time, which is the time for shocks from the surrounding hot wind to propagate through the cold cloud. A number of simulations have determined that the shock cooling time for the clouds in question is $\sim 100$ years \cite{2009ApJ...703..330C,McClure-Griffiths:2013awa}. On the other hand, the cloud crushing times for the hot wind hypothesized in \cite{McClure-Griffiths:2013awa} are substantially longer, $1-10$ Myr, depending on the density of the cloud, indicating minimal internal disruption and allowing for the assumption of stability when modelling the entrained clouds.

Furthermore, the ability to accelerate the entrained clouds to the velocities and galactic radii at which they are observed, provides further constraints on cloud disruption processes. In \cite{2017MNRAS.468.4801Z}, it is shown that simulations which neglect magnetic fields, produce cloud shredding timescales over which the clouds cannot be accelerated by galactic winds to their observed velocities. Magnetic fields are therefore likely to be critical in suppressing cloud disrupting processes such as evaporation and turbulent instabilities. The importance of internal magnetic fields in suppressing the disruption of entrained clouds was demonstrated in full magnetohydrodynamic simulations by \cite{2015MNRAS.449....2M}, while the suppression of hydrodynamic instabilities due to magnetic fields is discussed in \cite{2008ApJ...678..274O}. Therefore, because our bounds do not rely on the assumption of equilibrium, and moreover the literature indicates these clouds are stable even in the presence of the hypothetical hot wind, we think that it is reasonable to derive bounds using cold Galactic Center gas clouds.

\acknowledgments 

We wish to thank Navid Abbasi, Maxim Pospelov, Nirmal Raj, Farid Taghinavaz, and Aaron Vincent for discussions. J.~B.~and F.~E.~thank the CERN theory group for hospitality. J.~B.~thanks the organizers of the KITP High Energy Physics at the Sensitivity Frontier Workshop; this research was supported in part by the NSF under Grant No.~PHY-1748958. Research at Perimeter Institute is supported by the Government of Canada through Industry Canada and by the Province of Ontario through the Ministry of Economic Development \& Innovation. A.~B., J.~B., and S.~S.~ acknowledge the support of the Natural Sciences and Engineering Research Council of Canada. J.~B.~thanks the Aspen Center for Physics, which is supported by NSF Grant No.~PHY-1066293.

\bibliography{gcgcb}

\end{document}